\documentclass[aps,prx,twocolumn,groupedaddress]{revtex4-2}

\usepackage{hyperref}
\hypersetup{colorlinks=true, linkcolor=black, citecolor=black, urlcolor=black}

\usepackage{color}
\usepackage{xcolor}
\usepackage[dvipsnames]{xcolor}

\usepackage[utf8]{inputenc}
\usepackage[english]{babel}
\usepackage{amsmath,graphicx,enumerate}

\begin{document}


\title{Sparkling bubbles in chiral active fluids}

\author{A. Petrini$^{1}$}
\author{R. Maire$^{2}$}
\author{U. Marini Bettolo Marconi$^{3}$}
\author{L. Caprini$^{1}$}
\email{lorenzo.caprini@uniroma1.it}
\affiliation{$^1$ Physics Department, Sapienza University of Rome, Piazzale Aldo Moro 5, Rome, Italy }
\affiliation{$^2$ Department of Condensed Matter, University of Barcelona, 08028 Barcelona, Spain}
\affiliation{$^3$ Physics Department, University of Camerino, Via Madonna delle Carceri, Camerino, Italy }


\newcommand{\beq}{\begin{equation}}
\newcommand{\eeq}{\end{equation}}
\newcommand{\bea}{\begin{eqnarray}}   
\newcommand{\eea}{\end{eqnarray}}

\newcommand{\br}{{\bf r}}
\newcommand{\bu}{{\bf u}}
\newcommand{\bv}{{\bf v}}
\newcommand{\bR}{{\bf R}}
\newcommand{\bRz}{{\bf R}^0}
\newcommand{\bk}{{ \bf k}}
\newcommand{\bx}{{ \bf x}}
\newcommand{\vv}{{\bf v}}
\newcommand{\bn}{{\bf n}}
\newcommand{\mb}{{\bf m}}
\newcommand{\bq}{{\bf q}}
\newcommand{\bK}{{\bf K}}
\newcommand{\rb}{{\bar r}}
\newcommand{\rr}{{\bf r}}
\newcommand{\eb}{{\bf e}}

\newcommand{\RB}{R_{\text{B}}}
\newcommand{\kk}{\boldsymbol{\kappa}}
\newcommand{\greeketabold}{\boldsymbol{\eta}}
\newcommand{\xxi}{\boldsymbol{\xi}}
\newcommand{\cchi}{\boldsymbol{\chi}}
\newcommand{\bomega}{\boldsymbol{\Omega}}

\date{\today}


\setlength{\bibsep}{0pt}

\makeatletter
\renewcommand{\bibsection}{%
  \section*{References}%
}
\makeatother

\begin{abstract}

We study an inertial chiral active fluid, formed by repulsive particles that transfer angular momentum through odd interactions, i.e. transverse forces.
Chirality induces an inhomogeneous phase, consisting of rotating bubbles, whose formation is favored at an optimal packing fraction.
In this regime, we discover that bubbles may be dynamically unstable, breaking up and reforming in the steady state, thereby showing a spontaneous sparkling-like behavior reminiscent of supersaturated liquids.
Bubbles and sparkling bubbles are predicted by a coarse-grained hydrodynamic theory, revealing the intrinsic non-linearity of these collective phenomena, and call for experimental verifications in granular spinners or spinning colloids.
\end{abstract}

\maketitle

\section{Introduction}

From the effervescent dance of bubbles in a glass of carbonated water to the slow rise of vapor pockets in a boiling liquid, the sparkling behavior of fluids has long fascinated both scientists and everyday observers. In such equilibrium systems, bubbles form and evolve due to well-understood thermodynamic mechanisms -- typically gas-liquid coexistence and nucleation governed by surface tension and pressure differences. Once formed, these bubbles grow, coalesce, and detach, tracing trajectories determined by buoyancy and viscous drag. The physics of these bubbles remains rooted in equilibrium thermodynamics, where detailed balance and free energy govern their dynamics.

By contrast, non-equilibrium systems, such as active matter~\cite{marchetti2013hydrodynamics, elgeti2015physics, bechinger2016active,fodor2022irreversibility}, exhibit a much broader repertoire of collective dynamics, often showing density inhomogeneity and spontaneous phase separation~\cite{cates2015motility, fily2012athermal, digregorio2018full}.  Examples of these active systems, converting energy into systematic motion, range from biological microorganisms~\cite{drescher2009dancing,mathijssen2019oscillatory,pellicciotta2025wall} and synthetic microswimmers at the micron scale~\cite{buttinoni2013dynamical,bricard2013emergence,ginot2015nonequilibrium,ketzetzi2025active} to active granular particles~\cite{aranson2007swirling, kudrolli2010concentration, kumar2014flocking, koumakis2016mechanism, baconnier2022selective, lopez2023spin, antonov2025self} and animals at the macroscopic scale~\cite{cavagna2014bird}. 
While several physical or biological systems are characterized by a linear motion, in most of the cases, these systems possess a degree of chirality~\cite{lowen2016chirality, liebchen2022chiral,mecke2024emergent}, due to an intrinsic handedness, that allows them to move in circular or helical trajectories. This is the case of granular spinners~\cite{scholz2018rotating, vega2022diffusive, siebers2023exploiting, farhadi2018dynamics, caprini2024spontaneous}, chiral active colloids~\cite{ebbens2010self, kummel2013circular}, sperm~\cite{woolley2003motility} or bacteria moving close to surfaces~\cite{lauga2006swimming}. 

Chirality is often introduced as a constant angular velocity~\cite{lowen2016chirality}, which suppresses clustering~\cite{liao2018clustering,ma2022dynamical,bickmann2022analytical}, shifts the melting point~\cite{kuroda2025long}, and leads to rotating patterns~\cite{liebchen2017collective,zhang2020reconfigurable,kreienkamp2022clustering}, traveling waves~\cite{liebchen2016pattern,pisegna2025spinning}, as well as self-rotating crystallites~\cite{huang2020dynamical} and vortices~\cite{kruk2020traveling,liao2021emergent} with a characteristic correlation length~\cite{shee2024emergent}.
However, when two rotating objects interact in a fluid, they experience effective transverse forces generated by the flow field induced by the other particle, as observed in Volvox algae~\cite{drescher2009dancing,ishikawa2020stability}. Similar transverse forces also govern the motion of colliding chiral active granular particles that perform circular trajectories, such as air-driven spinners~\cite{lopez2022chirality}.
These transverse forces, often referred to as odd interactions~\cite{caprini2025modeling}, generically arise from chirality-induced translation–rotation coupling and control the dynamics of rotating objects.


Microscopic odd (transverse) interactions in chiral liquids with strong attractions have been recently introduced in Ref.~\cite{caporusso2024phase} to reproduce the edge currents~\cite{van2016spatiotemporal} experimentally observed in granular spinners~\cite{lopez2022chirality,caprini2025active} rotating colloids or starfish-embryos~\cite{tan2022odd} experiments at high density.
As shown theoretically and numerically, odd interactions naturally generate odd elasticity~\cite{scheibner2020odd, alexander2021layered, braverman2021topological, shankar2024active, kole2024chirality, lee2025odd} (see Ref.~\cite{fruchart2023odd} for a review), which manifests as a transverse response to an external perturbation.
Recently, these findings have been reproduced through a phase field crystal model giving rise to self-rotating crystallites~\cite{huang2025anomalous}.

However, it was recently discovered that an inertial ideal crystal subject to linear odd interactions can become linearly unstable at high chirality, i.e., for large transverse forces~\cite{caprini2025odd}. This instability manifests in high-density chiral active solids as a novel phase transition from a homogeneous to an inhomogeneous state characterized by bubbles induced by odd interactions, termed the BIO phase~\cite{caprini2025bubble}. This phenomenon is consistent with a coarse-grained hydrodynamic theory for inertial chiral particles, in which the linear instability of the homogeneous state arises from the interplay between a chirality-induced torque density and an odd viscosity~\cite{marconi2026emergent}. The BIO phase can thus be regarded as a general feature of inertial chiral systems: it has been observed in chiral rotors in a fluid, where transverse forces arise from hydrodynamic interactions~\cite{shen2023collective}, as well as in dry models of granular spinners, where odd interactions originate from tangential friction forces~\cite{digregorio2025phase}.

\begin{figure*}[!t]
\centering
\includegraphics[width=0.95\linewidth,keepaspectratio]{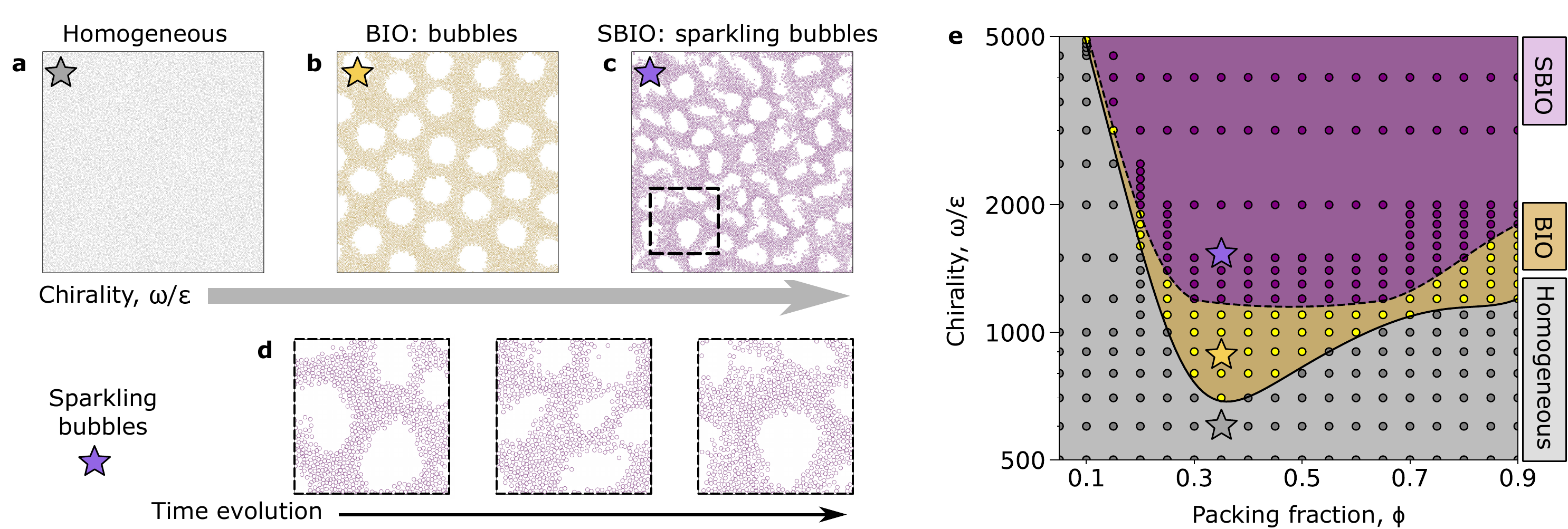}
\caption{
Sparkling bubbles induced by odd interactions (SBIO).
(a)–(c) Snapshot configurations of the homogeneous, bubble (BIO), and sparkling bubble (SBIO) phases, obtained at reduced chirality $\omega/\epsilon = 600, 900, 1500$ and packing fraction $\phi = 0.35$.
(d) Time evolution of a sparkling bubble (zoomed from (c)) at three different times for $\omega/\epsilon = 1500$ and $\phi = 0.35$.
(e) Phase diagram in the $(\omega/\epsilon, \phi)$ plane. In all panels, colors indicate the reference phase: homogeneous (gray), BIO (yellow), and SBIO (violet).
The remaining simulation parameters are $T_r = 20$ and $\tau_I=m/\gamma = 10^{-2}\tau$.
}\label{fig:Fig1_phasediagram}
\end{figure*}

Here, we show that chiral active fluids are not only characterized by the BIO phase, but also by a qualitatively distinct state that emerges at low density and high chirality. This phase, which we term a sparkling bubble phase (SBIO), is induced by chirality through odd interactions. In stark contrast to the BIO phase -- where bubbles remain approximately circular and maintain a stable size (Fig.~\ref{fig:Fig1_phasediagram}~(b)) -- the SBIO is marked by persistent dynamical activity: bubbles continuously nucleate, deform, break, and reform even in the steady state (Fig.~\ref{fig:Fig1_phasediagram}~(c) and Fig.~\ref{fig:Fig1_phasediagram}~(d)).
Unlike the bubbles in BIO, which remain stable, nearly circular, and almost immobile within a homogeneous, dense fluid of particles, these sparkling bubbles are dynamically unstable and continually regenerated by the non-equilibrium fluxes inherent in odd interactions. The resulting behavior is reminiscent of the effervescence observed in carbonated liquids, yet here it is sustained by chirality through transverse forces rather than by the release of dissolved gas.

Sparkling behavior in chiral systems has not been reported in previous studies. Indeed, chiral systems subject to hydrodynamic interactions have been shown to form bubbles (or cavities) with oscillatory size dynamics in the high-chirality regime~\cite{shen2023collective}. This behavior may be interpreted as a precursor to the sparkling bubble phenomenon observed here. In contrast, granular spinner models at high chirality exhibit a turbulent regime in which low- and high-density phases continuously evolve, with domains moving and rearranging over time. Whether this regime corresponds to the sparkling bubble state -- namely, a state characterized by individual bubbles with a well-defined average size -- or instead represents a phase-separated turbulent flow without persistent bubbles remains an open question requiring further investigation.

The sparkling bubble phase discovered here thus represents a novel mode of organization in chiral active fluids, which can be verified in experiments based on granular spinners or rotating colloids in a fluid.
This collective behavior bridges intuitive, everyday observations of bubbling in equilibrium fluids with the exotic, self-organized dynamics of matter far from equilibrium.

\section{Model}
We consider a two-dimensional chiral active fluid composed of $N$ repulsive particles subject to odd interactions, i.e., forces acting transversely to the line connecting particle pairs~\cite{caporusso2024phase,guo2025diffusion,caprini2025bubble}. Each particle is in contact with a thermal bath at temperature $T$ and evolves according to an underdamped equation of motion for its velocity, $\mathbf{\it v}_i=\dot{\mathbf{x}}_i$.
\begin{equation}
m \dot {\textbf{\textit{v}}}_i=- \gamma \textbf{\textit{v}}_i
+\mathbf{F}_i+\mathbf{F}^{\mathrm{odd}}_i + \sqrt{2\gamma k_B T} \,\boldsymbol{\eta}_i \,,
\label{eq:dynamics}
\end{equation}
where $\boldsymbol{\eta}_i$ is a white-noise term with zero mean and unit variance, $\gamma$ denotes the friction coefficient, and $k_B$ corresponds to the Boltzmann constant. The force ${\bf F}_i=-\nabla_i U^{\mathrm{tot}}$ derives from the total potential $U^{\mathrm{tot}}=\sum_{i<j}U(|\mathbf{x}_i-\mathbf{x}_j|)$, with $U(r)$ given by the purely repulsive Weeks–Chandler–Andersen potential~\cite{andersen1971relationship}, $U(r)=4\epsilon\big[(\sigma/r)^{12}-(\sigma/r)^{6}\big]+\epsilon$ for $r<2^{1/6}\sigma$ and $U(r)=0$ otherwise. Here $\sigma$ is the nominal particle diameter and $\epsilon$ sets the characteristic energy scale of the repulsive interactions. Chirality enters the dynamics through an additional interaction force, $\mathbf{F}_i^{\mathrm{odd}}$, referred to as the odd or transverse force, which can be written as
\begin{equation}
\mathbf{F}^{\mathrm{odd}}_i = - \nabla_i \sum_{k<j}  U^{\mathrm{odd}}(|\mathbf{x}_k-\mathbf{x}_j|)\times \hat{\mathbf{z}}\,,
\label{eq:odd_force}
\end{equation}
where $\hat{\mathbf{z}}$ is the unit vector normal to the plane of motion and $U^{\mathrm{odd}}(|\mathbf{x}_k-\mathbf{x}_j|)$ is a function of the interparticle distance. These forces act transversely to the line connecting particle centers due to the cross product with $\hat{\mathbf{z}}$: the $x$ component of $\mathbf{F}_i^{\mathrm{odd}}$ contributes to the $y$ dynamics, while the negative $y$ component contributes to the $x$ dynamics. Motivated by experimental studies of colloidal spinners governed by transverse interactions~\cite{massana2021arrested}, we choose a long-range form $U^{\mathrm{odd}}(r)=\omega(\sigma/r)$, cut and shifted at $r=5\sigma$, with $\omega$ setting the interaction strength. 
In what follows, we simply refer to $\omega$ as the chirality, since transverse forces control the dynamics of chiral objects rotating in a fluid or granular spinners subject to rotational friction~\cite{digregorio2025phase,caprini2025modeling}.
However, in the latter case, the long-range form of $U^{\mathrm{odd}}(r)$ should be replaced by a short-range interaction acting at contact.

Simulations of the dynamics~\eqref{eq:dynamics} are performed in a box of size $L$ with periodic boundary conditions. Lengths are rescaled by the particle diameter $\sigma$, and time by $\tau=\sigma\sqrt{m/\epsilon}$, while energy is measured in units of $\epsilon$. The system is characterized by three dimensionless parameters: the reduced chirality $\omega/\epsilon$, which measures the relative strength of transverse and repulsive interactions; the reduced inertial time $\tau_I/\tau$, where $\tau_I=m/\gamma$; and the reduced temperature $T_{r} = \frac{k_{B} T}{\epsilon}\frac{\tau}{\tau_{I}}$, which quantifies the noise strength relative to the potential energy scale. In this work, we focus on the roles of chirality $\omega/\epsilon$ and packing fraction $\phi = N\sigma^{2}\pi/(4L^{2})$, exploring values ranging from dilute gas-like to dense liquid-like conditions at equilibrium ($\omega=0$). The parameters $\tau_I/\tau$ and $T_r$ are kept fixed: the latter has a negligible effect, while the former merely shifts the location of the bubble transition previously reported~\cite{caprini2025bubble}. 
Inertia, however, is essential to generate an instability of the homogeneous phase and promote bubble formation~\cite{marconi2026emergent}. Consequently, $\tau_I/\tau$ must remain finite, as the overdamped limit of Eq.~\eqref{eq:dynamics} misses this mechanism.

\section{Results}

\subsection{Chirality-induced sparkling bubbles }

When the chirality $\omega$ is increased, the system undergoes a transition from a homogeneous phase (Fig.~\ref{fig:Fig1_phasediagram}(a) and Supplemental Video~1) to an inhomogeneous phase (Fig.~\ref{fig:Fig1_phasediagram}(b) and Supplemental Video~2), recently termed BIO, for bubbles induced by odd interactions. The BIO phase, first observed numerically in Ref.~\cite{caprini2025bubble}, is characterized by nearly circular bubbles, i.e., particle-free regions corresponding to stable cavitated areas within the fluid.

In contrast to previous studies, we find that increasing $\omega$ at low packing fraction destabilizes the BIO phase and gives rise to an unprecedented dynamically unstable state. In this regime, bubbles acquire irregular shapes with large oscillations (Fig.~\ref{fig:Fig1_phasediagram}(c)), until they continuously collapse and dynamically reform (Fig.~\ref{fig:Fig1_phasediagram}(d) and Supplemental Video~3). This phase will be referred to as the sparkling bubble phase (SBIO), in analogy with the sparkling behavior observed, for example, in carbonated liquids. 
However, in such supersaturated (metastable) systems, bubbles nucleate via thermal fluctuations that overcome the free-energy barrier imposed by surface tension. In contrast, the sparkling bubbles observed in our study emerge spontaneously as a consequence of odd interactions and the intrinsically nonequilibrium nature of the chiral active fluid.
Although both BIO and SBIO are characterized by the presence of bubbles, they differ fundamentally in their dynamical behavior: in the BIO phase, bubbles are stable and maintain a finite size, whereas in the SBIO phase, they are dynamically unstable and exhibit large size fluctuations.

In addition, we discover that bubble phases do not require a high-density or solid-like state to emerge but already appear in low-density chiral active fluids, even at small packing fractions ($\phi \gtrsim 0.1$). This contrasts with Ref.~\cite{caprini2025bubble}, where bubbles were observed only in solid-like states. Our phase diagram in the plane of chirality $\omega/\epsilon$ and packing fraction $\phi$ (Fig.~\ref{fig:Fig1_phasediagram}(e)) displays a U-shaped convex structure, indicating the existence of an optimal, intermediate $\phi$ at which bubble formation is most favored and the sparkling bubble phase appears. 
This U-shaped phase diagram is consistent with recent results on chiral granular spinners, which generate effective odd interactions through tangential friction~\cite{digregorio2025phase}.

The U-shape profile of the phase diagram with chirality and packing fraction can be intuitively understood.
Indeed, bubbles require larger chirality to nucleate both at low and high densities: In the dilute limit, transverse forces become ineffective (as if $\omega=0$), and the system behaves as an equilibrium ideal gas of repulsive particles showing a homogeneous state. At large densities, by contrast, the pressure generated by strong repulsive interactions increases, so a higher chirality is needed to induce a density inhomogeneity~\cite{marconi2026emergent}.


\subsection{Vorticity and bubble size}

\begin{figure}[!t]
\centering
\includegraphics[width=1\linewidth,keepaspectratio]{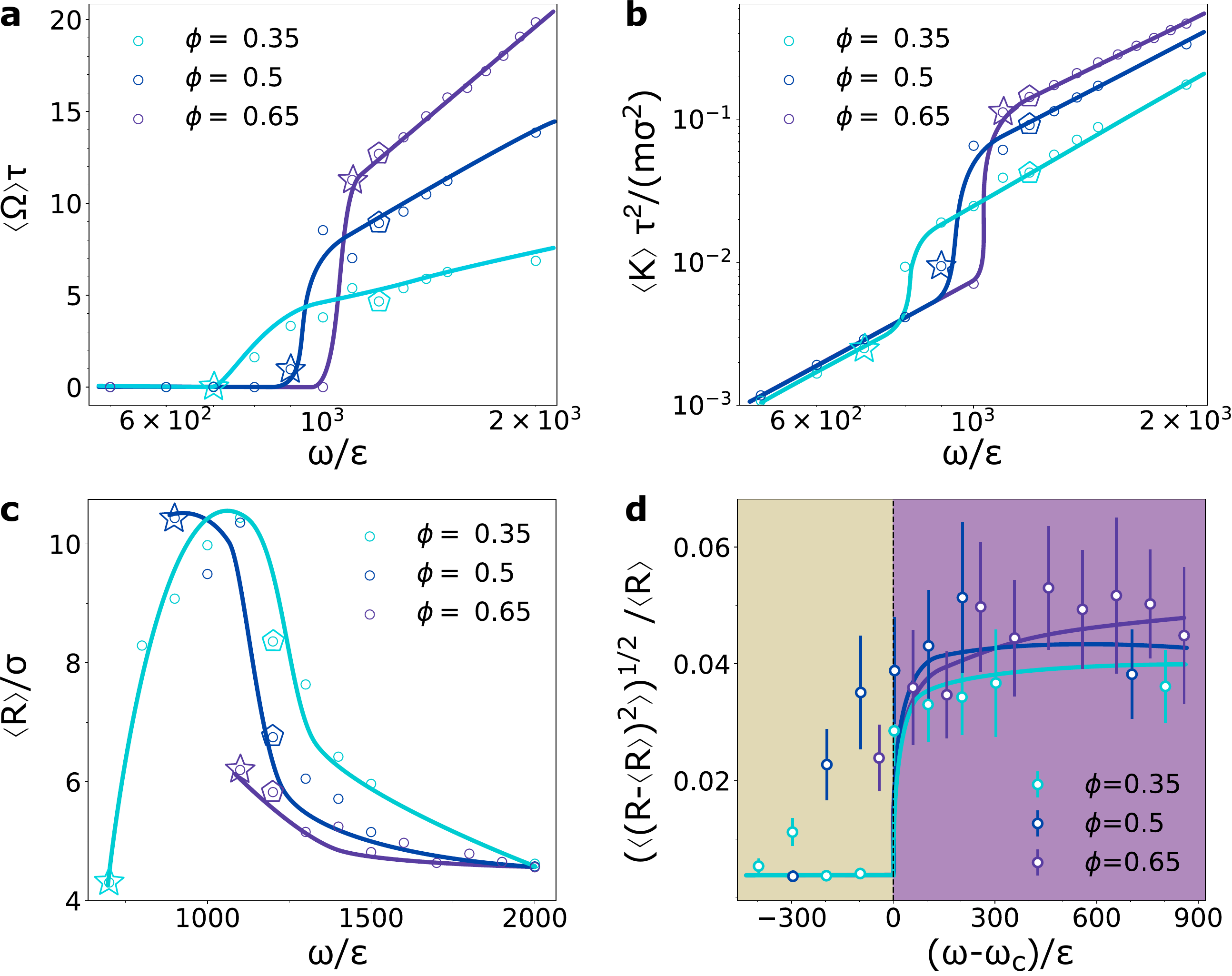}
\caption{
BIO and SBIO phases.
(a) Average vorticity $\langle \Omega \rangle$, (b) average kinetic energy $\langle K \rangle$, and (c) average bubble radius $\langle R \rangle$ as functions of the chirality $\omega$ (rescaled by $\tau = \sigma \sqrt{m/\epsilon}$) for three packing fractions, $\phi = 0.35, 0.5, 0.65$.
(d) Square root of the normalized variance of the bubble size, $\frac{\langle R^2\rangle-\langle R\rangle^2}{\langle R\rangle^2}$, as a function of $\omega - \omega_c$. Here, $\omega_c$ denotes the critical chirality generating the BIO–SBIO transition.
Points are obtained from simulations, while solid lines serve as guides to the eye. 
Stars and pentagons in panels (a), (b), and (c) correspond to the first points (at fixed $\phi$) at which the BIO or SBIO phase is observed.
The remaining simulation parameters are $T_r = 20$ and $\tau_I=m/\gamma = 10^{-2}\tau$. }\label{fig:Fig2_averages}
\end{figure}

\begin{figure*}[!t]
\centering
\includegraphics[width=0.8\linewidth,keepaspectratio]{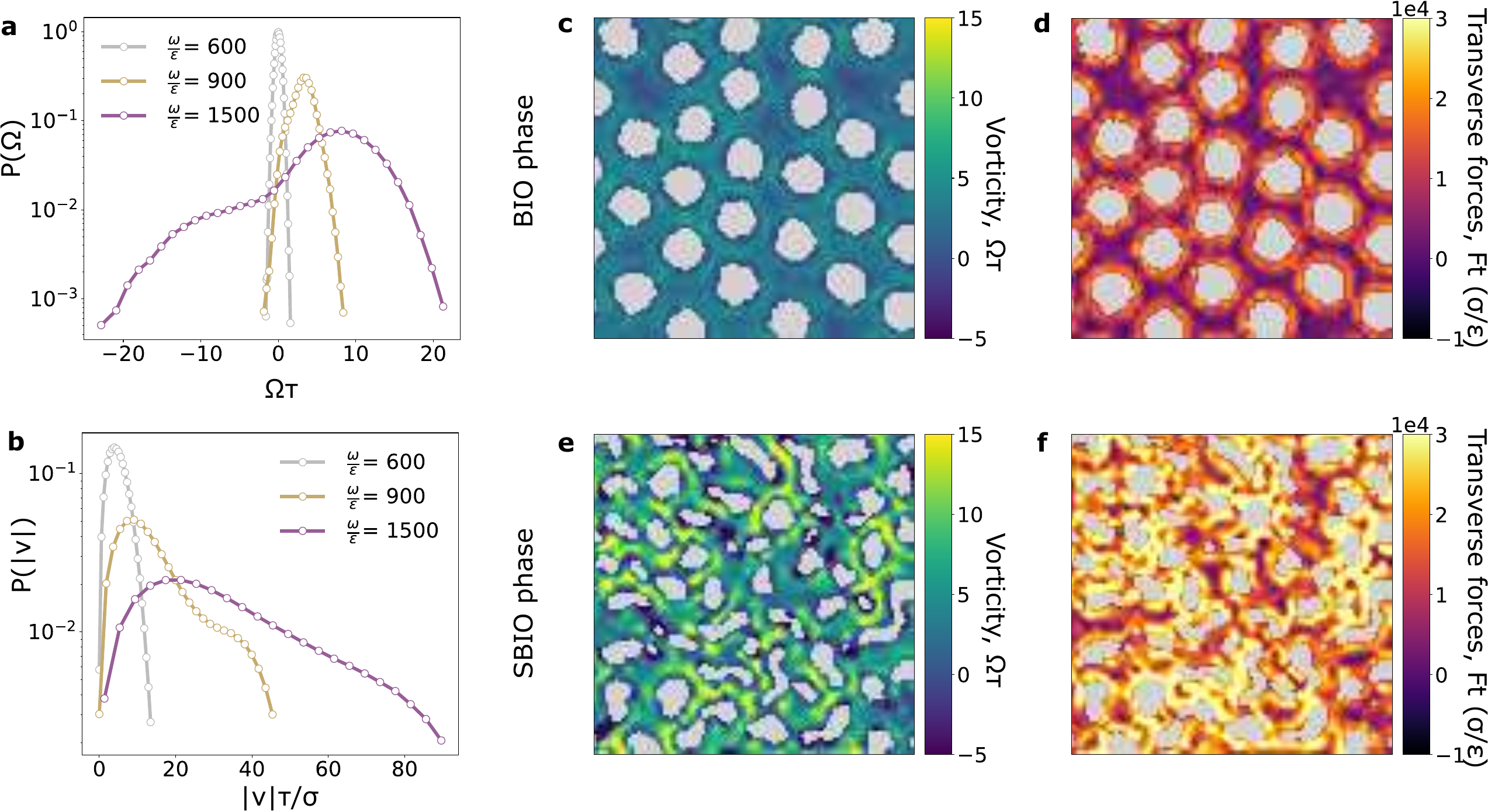}
\caption{
Vorticity induced by sparkling behavior.
(a) Vorticity distribution $P(\Omega)$ and (b) speed distribution $P(|\mathbf{v}|)$ for several values of the chirality $\omega$ (rescaled by $\epsilon$), spanning the homogeneous, BIO, and SBIO phases.
(c)-(f) Snapshots of the system showing the magnitude of the vorticity ((c) and (e)) and the magnitude of the transverse force ((d) and (f)) as a color gradient. Gray regions indicate areas devoid of particles. Panels (c) and (d) correspond to a BIO configuration ($\omega/\epsilon = 900$), while panels (e) and (f) correspond to a SBIO configuration ($\omega/\epsilon = 1500$). 
The remaining simulation parameters are $\phi = 0.35$, $T_r = 20$, and $\tau_I=m/\gamma = 10^{-2}\tau$.
}\label{fig:Fig3_averages}
\end{figure*}

The homogeneous and BIO phases can be distinguished by monitoring the steady-state average vorticity, $\langle \Omega \rangle$ (Fig.~\ref{fig:Fig2_averages}(a)), as a function of chirality $\omega$. 
The vorticity field is defined as $\Omega(\mathbf{x}) = \partial_x u_y(\mathbf{x}) - \partial_y u_x(\mathbf{x})$, where $u_x(\mathbf{x})$ and $u_y(\mathbf{x})$ are the Cartesian components of the velocity field. 
In a homogeneous configuration, $\langle \Omega \rangle$ is close to zero, whereas it jumps to a high value in the BIO phase, independently of the packing fraction (see also the plot of $\Omega(\mathbf{x})$, Fig.~\ref{fig:Fig3_averages}(c)).
This local vorticity is a consequence of the edge currents observed at the bubble interface~\cite{caprini2025bubble}, where transverse forces are unbalanced (heatmap in Fig.~\ref{fig:Fig3_averages}(d)) and lead the particles to rotate at high speed. 
The resulting local velocity increase leads to a sharp rise in the average kinetic energy per particle, $\langle K \rangle$, at the homogeneous-to-BIO transition (Fig.~\ref{fig:Fig2_averages}(b)), where $K$ is defined as $K = m \sum_{i=1}^{N} \mathbf{v}_i^2/(2N)$.
As the system approaches the sparkling bubble phase (SBIO), both $\langle \Omega \rangle$ and $\langle K \rangle$ increase smoothly and monotonically with chirality $\omega$ (Fig.~\ref{fig:Fig2_averages}(a)-(b)). 
However, in the SBIO, the vorticity distribution $P(\Omega)$ is not only peaked at high $\Omega > 0$ but also develops a long tail for $\Omega < 0$, indicating the simultaneous presence of vortices and antivortices (Fig.~\ref{fig:Fig3_averages}(a)), which are absent in the BIO phase. This results in richer structures of $\Omega(\mathbf{x})$ localized in the bulk (heatmap in Fig.~\ref{fig:Fig3_averages}(e)), which evolve dynamically due to unbalanced odd interactions even in the bulk (heatmap in Fig.~\ref{fig:Fig3_averages}(f)). These interactions generate higher kinetic energy and a broad speed distribution $P(|\mathbf{v}|)$ with a heavy tail (Fig.~\ref{fig:Fig3_averages}(b)).

To characterize the BIO–SBIO transition, we focus on the properties of the bubbles. The average bubble radius, $\langle R \rangle$, increases monotonically with $\omega$ in the BIO phase (Fig.~\ref{fig:Fig2_averages}(c)). Upon entering the SBIO phase, however, $\langle R \rangle$ decreases with increasing $\omega$, indicating that sparkling bubbles are smaller than the stable bubbles of the BIO phase. To quantitatively distinguish between BIO and SBIO, we consider the normalized standard deviation of the bubble size, $\sqrt{\langle (R-\langle R\rangle)^2 \rangle} / \langle R \rangle$, as an order parameter of the sparkling behavior numerically observed. This quantity is evaluated for different packing fractions $\phi$ as a function of $\omega - \omega_c$, where $\omega_c$ denotes the chirality at which the BIO–SBIO transition occurs. The normalized variance exhibits a sharp jump from low to high values, confirming that the SBIO is characterized by bubbles with strongly fluctuating sizes (Fig.~\ref{fig:Fig2_averages}(d)). Just before the transition to the SBIO phase, the stable bubbles of the BIO phase begin to exhibit pronounced radial oscillations while remaining spatially ordered. This observation motivates the theory developed below.

\subsection{Theory: Generalized Rayleigh-Plesset approach}

The stability of bubbles can be understood through a mechanical argument. While repulsive interactions generate the usual pressure that tends to suppress bubble formation, odd interactions produce edge currents~\cite{caprini2025bubble} at the bubble interface. Rotating particles are additionally subject to a centrifugal force directed outward relative to the bubble surface, which balances the pressure and makes bubbles stable.
This mechanism can be formalized through a hydrodynamic theory derived from the dynamics~\eqref{eq:dynamics}, yielding a continuity equation for the density $\rho$ and a momentum equation for the velocity field $\mathbf{u}$ (See the Methods). By expressing the velocity in polar coordinates centered on a bubble, we obtain equations for the radial and tangential components, $u_r$ and $u_\theta$, where $r$ is the distance from the bubble center and $\theta$ the azimuthal angle. Evaluating these equations in the steady state, where spatial gradients of the velocity field can be neglected, we find: 
\begin{subequations}
\label{eq:hydro_eq_final}
\begin{align}
\label{eq:hydro_theta}
&\gamma u_\theta = f_\theta^{\mathrm{odd}} \\
\label{eq:hydro_radial}
&\frac{u_\theta^2}{r} = \frac{\partial_r P}{\rho}\,,
\end{align}
\end{subequations}
where $P$ is the pressure, which includes the repulsive WCA interactions and $f_\theta^{\mathrm{odd}}$ is the tangential component of the average odd force (see Methods for their definition). 
We note that $f_\theta^{\mathrm{odd}}$ is uniquely generated by chirality through odd interactions and is related to the torque density that appears at first order in the density gradients in the hydrodynamic treatment reported in Ref.~\cite{marconi2026emergent}.

Equation~\eqref{eq:hydro_theta} expresses the force balance along the tangential direction relative to the bubble center and predicts a non-vanishing steady-state tangential velocity, $u_\theta$, generated by transverse forces, which produce a non-zero tangential force $f_\theta^{\mathrm{odd}}$ around the bubble. In passive fluids, $f_\theta^{\rm odd} \equiv 0$ because transverse chiral forces are absent, and therefore $u_\theta = 0$. In chiral fluids, $f_\theta^{\rm odd}$ cancels in the bulk due to local force balance among neighboring particles. At the bubble interface, this balance is broken, as the dilute interior cannot compensate for the transverse force exerted by the dense phase. The remaining interfacial $f_\theta^{\rm odd}$ is precisely what drives the localized edge currents observed numerically.
Eq.~\eqref{eq:hydro_radial} describes the radial force balance, involving the pressure gradient and repulsive forces, which tend to increase the pressure. 
Here, however, the nonzero $u_\theta$ generates a centrifugal term, $u_\theta^2/r$, directed outward from the bubble center. For high chirality, $f_\theta^{\mathrm{odd}}$ and, thus, $u_\theta$ are large and can balance the pressure contributions, creating a density inhomogeneity, i.e., a bubble.

To qualitatively understand the sparkling bubble phase, we generalize to a chiral active fluid the method derived by Rayleigh and Plesset to describe bubble evolution in equilibrium~\cite{rayleigh1917pressure, plesset1949dynamics}. Mass conservation implies that $2\pi r\, u_r(r,t)\,\Delta t = \dot{A}\,\Delta t$, where $A$ is the bubble area. For a circular bubble, the area change entails a change of the bubble radius $R$, which sets a relation between the radial fluid velocity at the interface and the bubble growth rate, $u_r(r,t) = \dot{R}\, R/r$. By replacing $u_r$ with $\dot R$ in the momentum equation evaluated at $r=R$, we obtain:
 \begin{equation}
 \label{eq:R}
  	\ddot R +\frac{\gamma}{m} \dot R-  \frac{\bar{u}_\theta^2}{R} =  -\frac{1}{\rho}\left(\frac{\partial P}{\partial r}\right)_{r=R} \,,
  \end{equation}
  with $\bar{u}_\theta = u_\theta(R)$ and $\rho=\rho(R)$. 
 We solve Eq.~\eqref{eq:R} perturbatively by considering a linear perturbation around its steady-state solution, i.e., $R(t) = \RB \bigl(1 + x(t)\bigr)$, where $\RB$ is the steady-state bubble radius and $x(t)$ is a small dimensionless perturbation, which is predicted to evolve as:
\begin{equation}
	\RB x(t)= A  e^{- \frac{\gamma t}{2m}}  \sin{\left(\mu t + \psi\right)} \,,
\end{equation}
where $A$ and $\psi$ are two constants (see Methods). Here, $\mu$ is the Minnaert frequency associated with the bubble oscillation, which reads
\begin{equation}
\label{eq:omega}
	\mu= \sqrt{	 \dfrac{\bar{u}^2_\theta }{\RB^2}  - \dfrac{\gamma^2}{4m^{2}}} \,. 
\end{equation}
Expression~\eqref{eq:omega} implies that the radius relaxes exponentially to its steady-state for small $\bar{u}_\theta(\RB)$, i.e., low chirality $\omega$, while for sufficiently large $\bar{u}_\theta$ the exponential relaxation is modulated by time oscillations.

This oscillatory behavior of the bubble radius provides an initial indication of the mechanisms underlying the sparkling dynamics observed at high chirality, but it should not be interpreted as a theoretical prediction of the BIO–SBIO transition. Indeed, our calculations show that high chirality promotes strong deformations of the bubble size, leading to continuous stretching and compression over time. When such deformations occur, neighboring bubbles can interact and interfere with one another, giving rise to effective interactions that destabilize the bubbles and may result in merging or fragmentation.
However, our current theoretical framework does not capture this instability. One possible reason for this discrepancy is that the theory does not include viscous and odd-viscous contributions in the momentum equation. In addition, Eq.~\eqref{eq:R} assumes perfectly circular bubbles, whereas shape deformations may occur and constitute a further source of instability. Incorporating these effects remains an important direction for future work.



\section{Conclusions}

 Here, we have shown that chiral active fluids exhibit sparkling bubbles (SBIO), a dynamical state reminiscent of carbonation but arising without any supersaturation. Compared to the bubble (BIO) phase~\cite{caprini2025bubble}, in which bubbles are circular and stable, the SBIO phase is characterized by unstable bubbles that spontaneously form, break, and reform in the steady state. These qualitative differences in dynamics indicate that SBIO constitutes a distinct phase from BIO. We further note that this sparkling phase emerges in low-density systems, where bubbles are more readily observed, as evidenced by a U-shaped phase diagram in the plane of packing fraction and chirality. This density dependence is consistent with phase diagrams reported in models of chiral granular spinners, where odd interactions arise from tangential friction during collisions.

The spontaneous formation, breakup, and reformation of bubble-like cavities demonstrate that chirality alone can drive interfacial dynamics with no equilibrium analogue. This mechanism organizes microscopic chiral or driven processes into structures typically associated with gas–liquid imbalances, yet here generated purely by nonequilibrium chirality. The resulting sparkling state highlights a qualitatively richer landscape of dynamical phases in chiral active matter, potentially connected to odd viscosity~\cite{banerjee2017odd, markovich2021odd, reichhardt2022active, lou2022odd, hosaka2023lorentz, machado2023hamiltonian, markovich2021odd,markovich2025chiral}, odd mobility~\cite{poggioli2023odd, hargus2025odd}, and odd diffusivity~\cite{hargus2021odd, kalz2022collisions, vega2022diffusive, kalz2024oscillatory} that typify chiral systems.


\section{Methods}

\subsection{Simulation details}
To simulate the dynamics, it is convenient to recast the equations of motion in a non-dimensional form. We rescale lengths by the particle diameter $\sigma$, times by the characteristic dynamical time $\tau = \sigma\sqrt{m/\epsilon}$, and interaction strengths by the typical energy scale $\epsilon$. Introducing the inertial time $\tau_{I} = m / \gamma$, Eq. \eqref{eq:dynamics} can be rewritten as an equation for the dimensionless velocity $\textbf{v}'_i=\textbf{v}_i\tau/\sigma$:
\begin{equation}
 \dot {\textbf{\textit{v}}'}_i = -\dfrac{\tau}{\tau_{I}}\textbf{\textit{v}}'_{i}
+\mathbf{f}_{i} 
+ \dfrac{\omega}{\epsilon}\mathbf{f}_{i}^{\mathrm{odd}}
+ \sqrt{2T_{r}}\,\boldsymbol{\eta}'_{i}\,,
\label{eq:dynamics_nondim}
\end{equation}
where $\boldsymbol{\eta}'=\boldsymbol{\eta}\sqrt{\tau}$ is the dimensionless thermal noise and we have introduced the symbols $\mathbf{f}_{i}$ and $\mathbf{f}_{i}^{\mathrm{odd}}$ as dimensionless forces such that $\mathbf{F}_i = \mathbf{f}_{i} \epsilon/\sigma $ and $\mathbf{F}_i^{\mathrm{odd}} =  \mathbf{f}_{i}^{\mathrm{odd}}\omega/\sigma$.
In addition, we have defined the reduced temperature $T_{r} = \frac{k_{B} T}{\epsilon}\frac{\tau}{\tau_{I}}$ and the non-dimensional chirality parameter $\omega/\epsilon$. 
%
We numerically integrate Eq.~\eqref{eq:dynamics_nondim} for a system of $N = 10^{4}$ particles, fixing $\tau/\tau_{I} = 100$ and $T_{r} = 20$, while varying $\omega/\epsilon$ and the packing fraction $\phi$. 
The system is initialized from a configuration of non-overlapping particles uniformly distributed in a square simulation box of side $L$ with periodic boundary conditions (PBCs). Time evolution is performed using a stochastic Euler integration scheme. Interactions are handled through a \textit{Cell List} algorithm, with a cutoff of $5\sigma$, which does not qualitatively change the physics of the system~\cite{caprini2025bubble}.
Fields are then reconstructed via a numerical coarse-graining procedure, while bubble identification relies on an algorithm inspired by multi-spin clustering methods~\cite{Newman1999MonteCarloMethods}. 

The bubble radius $R$ in Fig.~\ref{fig:Fig2_averages} provides a measure of the average bubble size and does not account for the bubble shape. It is computed as follows: (i) identifying the bubble area, defined as the connected regions of low density; (ii) taking the square root of this area divided by $\pi$. In this way, we obtain the effective radius corresponding to a circle with the same area as the bubble.

\subsection{Derivation of Hydrodynamics}

In this section, we derive the predictions~\eqref{eq:hydro_eq_final} to elucidate the mechanism underlying bubble formation. Starting from the microscopic Langevin description, we derive the coarse-grained Kramers–Fokker–Planck equation for the joint distribution function of the system, $f_{N} = f_{N}({\rr,\vv},t)$, where the notation ${\rr, \vv}$ denotes the collection of $4N$-dimensional dynamical variables. By integrating out the positions and velocities of $(N-1)$ particles from the Kramers–Fokker–Planck equation, one obtains the evolution equation for the single-particle distribution function, $f_{1} = f_{1}(\rr,\vv,t)$, which involves the two-particle distribution function $f_{2} = f_{2}(\rr, \rr^{,\prime}, \vv, \vv^{,\prime}, t)$. To close the hierarchy, we employ a two-body approximation, expressing $f_{2}$ in terms of the single-particle distribution function and the pair-correlation function $g_{2} = g_{2}\left(|\rr-\rr^{,\prime}|\right)$. Following this procedure~\cite{te2023derive}, we arrive at the following closed kinetic Boltzmann equation for the single-particle distribution function:
%
\begin{equation}
\begin{split}
    & \left( \partial_{t} + \vv \cdot \mathbf{\nabla}_{\rr} + \dfrac{1}{m}\int d\rr^{\,\prime}d\vv^{\,\prime}f_{1}^{\,\prime}g_{2} \left( \mathbf{F} + \mathbf{F}^{\mathrm{odd}}\right)\cdot \mathbf{\nabla}_{\vv} \right)f_{1} \\
    &= \dfrac{\gamma}{m}\mathbf{\nabla}_{\vv} \cdot \left( \vv + \dfrac{k_B T}{m}\mathbf{\nabla}_{\vv} \right) f_{1}  \,,
\end{split}
\label{eq:FPE_SingleParticle}
\end{equation} 
where $f_{1}^{\,\prime} = f_{1}(\rr^{\,\prime}, \vv^{\,\prime}, t)$. We define the hydrodynamic fields $\rho(\rr, t)$, $\mathbf{u}(\rr, t)$, and $\tilde{\mathbf{P}}(\rr, t)$, namely the mass density, velocity, and kinetic pressure tensor fields, as velocity moments of $f_{1}$:
\begin{equation}
\begin{pmatrix}
\rho \\
\rho\,\mathbf{u} \\
\tilde{\mathbf{P}} 
\end{pmatrix} = \int d\vv 
\begin{pmatrix}
m \\
m\vv \\
m  (\vv - \mathbf u) \otimes (\vv - \mathbf u)  
\end{pmatrix} f_{1}\,,
\label{eq:hydro_fields}
\end{equation}
where the $\otimes$ denotes the dyadic product. The hydrodynamic equations for the fields defined in Eq.~\eqref{eq:hydro_fields} are obtained by multiplying the kinetic Boltzmann equation (Eq.~\eqref{eq:FPE_SingleParticle}) by $\{m,\, m\vv,\, m  \vv \otimes \vv \}$ and integrating over the velocity. We close the hydrodynamic hierarchy at the level of the kinetic pressure tensor by imposing $\tilde{\mathbf{P}}_{i,j} = \delta_{i,j} k_B T\rho/m$. The hydrodynamic equations for the two slowest hydrodynamic fields then read:
\begin{eqnarray}
&&\partial_{t}\rho = - \mathbf{\nabla}\cdot\left(\rho\mathbf{u}\,\right) 
\label{eq:bcontinuity}\\
&& \partial_{t}\mathbf{u} + (\mathbf{u}\cdot\boldsymbol{\nabla})\mathbf{u} = 
- \dfrac{1}{\rho} \mathbf{\nabla} P- \frac{\gamma}{m}\mathbf{u} +\dfrac{\mathbf{f}^{\mathrm{odd}}}{m}\,,
\label{eq:momentumeq}
\end{eqnarray}
where we have neglected the viscous term. In Eq.~\eqref{eq:momentumeq} we have also defined the pressure gradient and the odd force as:
\begin{subequations}
\begin{align}
\mathbf{\nabla} P(\rr, t) &= \dfrac{k_B T}{m} \mathbf{\nabla} \rho+ \dfrac{\rho}{m^2}\int d\rr^{\,\prime} \rho^{\,\prime} g_{2}\mathbf{F}\left(\rr-\rr^{\,\prime}\right) 
\label{eq:gradP}\\
\mathbf{f}^{\mathrm{odd}}(\rr, t) &=\int d\rr^{\,\prime}\dfrac{\rho^{\,\prime}}{m}g_{2}\mathbf{F}^{\mathrm{odd}}\left(\rr-\rr^{\,\prime}\right)\,.
\label{eq:f_odd}
\end{align}
\end{subequations}
Eq.~\eqref{eq:momentumeq} differs from the inviscid Navier–Stokes equation of an ordinary fluid due to the presence of an additional average odd force term ${\bf f}^{\mathrm{odd}}$. This term corresponds to the average of odd interactions that break the mirror symmetry of the system, as evidenced by the expression of $\mathbf{F}^{\mathrm{odd}}$ in Eq.~\eqref{eq:f_odd}.

To elucidate the dynamical role of this additional term in the hydrodynamic equations, we assume the system to be in a phase-space configuration belonging to the BIO phase.
The system then develops a nearly circular cavity around a randomly selected pinning point. We therefore choose a reference frame centered on the forming bubble and adopt polar coordinates $(r,\theta)$, decomposing the velocity field into its radial and tangential components $(u_r, u_\theta)$. In this framework, Eqs.~\eqref{eq:bcontinuity} and \eqref{eq:momentumeq} take the form:
\begin{subequations}
\begin{flalign}
\label{eq:radial_Density}
&\frac{\partial \rho}{\partial t}+\frac{\partial_r (\rho r u_r)}{r} +\frac{\partial_\theta (\rho u_\theta)}{r}  =\, 0 \\
\label{eq:radial_NS}
&\!\left(\!\partial_t  + u_r\partial_r + \frac{u_\theta}{r} \partial_\theta + \frac{\gamma}{m} \right) u_r-\frac{u_\theta^2}{r} =
-\dfrac{\partial_r P}{\rho} \\
\label{eq:polar_NS}
&\!\left(\!\partial_t  + u_r\partial_r + \frac{u_\theta}{r} \partial_\theta + \frac{\gamma}{m} + \frac{u_r}{r}\!\right)\! u_\theta = \frac{f^{\mathrm{odd}}_{\theta}}{m} - \frac{\partial_\theta P}{r\rho} 
\end{flalign}
\end{subequations}
where $f_\theta^{\mathrm{odd}}$ corresponds to the non-vanishing tangential component of the average odd force.
By assuming rotational symmetry and stationarity, i.e.\ neglecting time derivatives and spatial gradients, we recover Eqs.~\eqref{eq:hydro_theta} and~\eqref{eq:hydro_radial}.

\subsection{Rayleigh-Plesset equation}

Let us assume that a circular cavity of radius $R(t)$ and area $A=\pi R^2$ is located at the origin of the coordinate system for a fluid of constant density $\rho_{0}$ and hosts a constant edge current circulating along its boundary. We consider the evolution equation for the normal component of the fluid velocity $u_r(r,t)$ around the cavity, as given by Eq.~\eqref{eq:radial_NS}.
We assume the fluid to be nearly incompressible, i.e. $\frac{1}{r}\partial_r(r u_r(r))=0$. Under these assumptions, the radial velocity field of the fluid at the cavity boundary coincides with the temporal variation of the cavity radius, $\dot R(t)$.
Using mass conservation, we impose a relation between the rate of change of the area occupied by the bubble, $\dot A$, and the particle flux normal to the circular interface:
\begin{equation}
	\dot A \rho_{0} \Delta t= 2 \pi r  u_r(r,t) \rho_{0} \Delta t
\end{equation}
which establishes a closed relation between the radial velocity field and $\dot R(t)$:
\begin{equation}
\label{eq:radial_velocity_RP}
 	u_r(r,t) = \dfrac{R}{r} \dot R\,.
\end{equation}
By replacing Eq.~\eqref{eq:radial_velocity_RP} in Eq.~\eqref{eq:radial_NS}, we obtain:
\begin{equation}
  \dfrac{ R \ddot R}{r}+\dfrac{ \dot R^2}{r}-\dfrac{ R^2 \dot R^2}{r^3}= \frac{u_\theta^2}{r} -\frac{1}{\rho}\left(\dfrac{\partial P}{\partial r}\right) -\dfrac{\gamma }{m} \frac{ R}{r}\dot R\,.
\label{eq:radiusequation}
\end{equation}
We now seek a solution of Eq.~\eqref{eq:radiusequation} evaluated at $r=R$. Moreover, we chose as equilibrium radius the bubble radius $\RB$ and we linearize the resulting equation around small deviations from the steady configuration of the bubble, i.e. by expanding around $R=\RB(1+ x)$. Recalling the condition of dynamical equilibrium,
$\frac{u_\theta^2}{\RB} =  \frac{1}{\rho}(\frac{\partial P}{\partial r})_{r=\RB}$, we find that the deviations form the equilibrium configuration, $x(t)$,  obey the equation
\begin{equation}
 \label{eq:linearized_RP}
	\ddot x +\dfrac{\gamma}{m} \dot x+  \dfrac{\bar{u}_\theta^2}{\RB^2} x = 0\,,
 \end{equation}
where $\bar{u}_{\theta} = u_{\theta}(\RB)$. Equation \eqref{eq:linearized_RP} corresponds to a damped harmonic oscillator with restoring term $\frac{\bar{u}_\theta^2}{\RB^2} x$. The general solution is a linear combination of exponentials characterized by the eigenvalues
 \begin{equation}
	\lambda_\pm=-\dfrac{\gamma}{2m}\pm \sqrt{\dfrac{\gamma^2}{4m^{2}}-\dfrac{\bar{u}_\theta^2}{\RB^2}}\,.
\end{equation}
This solution admits different dynamical regimes. In particular, as the tangential velocity increases, and specifically when $\frac{\bar{u}_\theta^2}{\RB^2} > \frac{\gamma^2}{4m^{2}}$, the system undergoes a transition from an overdamped to an underdamped regime. In the latter case, the bubble exhibits oscillations around the equilibrium radius $\RB$ with a characteristic Minnaert frequency~\cite{devaud2008minnaert}.
\begin{equation}
\mu = \sqrt{\dfrac{\bar{u}_\theta^2}{\RB^2}- \dfrac{\gamma^2}{4m^{2}}}\,.
\end{equation}
Let us consider a system displaying an initial perturbation such that $x(0)=D/\RB$ and $\dot x(0)=0$. The solution can then be written as
 \begin{equation}
 	\RB x(t) = Ae^{-\frac{\gamma}{2m}t}\sin(\mu t + \psi) \,,
\label{eq:solutionRP}
\end{equation}
where $A$ and $\psi$ are two constants, representing the oscillation amplitude and phase, with expressions:
\begin{subequations}
\begin{align}
&A= D\sqrt{1+\left(\dfrac{\gamma}{2m\mu}\right)^{2}}
\label{eq:Amplitude}\\
&\tan\left({\psi}\right) = \dfrac{2m\mu}{\gamma} = \sqrt{\dfrac{4m^{2}}{\gamma^{2}}\dfrac{\bar{u}_{\theta}^{2}}{\RB^{2}}-1}\,.
\label{eq:psi} 
\end{align}
\end{subequations}
The solution~\eqref{eq:solutionRP} provides a qualitative explanation for the continuous reshaping of bubbles in the SBIO phase: the radius oscillates over time, causing the domain to alternately stretch and compress until it encounters a neighboring bubble, leading to either merging or mutual annihilation.

\subsection{Description of the Supplemental Videos}

Supplemental Videos are obtained by integrating the dynamics in Eq.~\eqref{eq:dynamics} until the system reaches a steady state. Specifically, Supplemental Video~1 shows a homogeneous configuration at reduced chirality $\omega/\epsilon = 600$, corresponding to Fig.~\ref{fig:Fig1_phasediagram}(a). Supplemental Video~2 illustrates a typical time evolution from a homogeneous state to a bubble (BIO) phase at $\omega/\epsilon = 900$, corresponding to Fig.~\ref{fig:Fig1_phasediagram}(b). Finally, Supplemental Video~3 shows the evolution of a sparkling bubble (SBIO) phase at $\omega/\epsilon = 1500$, corresponding to Fig.~\ref{fig:Fig1_phasediagram}(c).
The remaining simulation parameters are $\phi = 0.35$, $T_r = 20$, and $m/\gamma = 10^{-2}\tau$.


\section*{Data availability}
\noindent
The data that support the plots within this paper and other findings of this study are available from the corresponding author upon request, while the Supplemental Videos are uploaded as Supplemental Material.

\section*{Code availability}
\noindent
The code to generate data, by using numerical simulations, is available under reasonable requests.

\section*{Acknowledgements}
\noindent
L.C. acknowledges support from the Italian Ministero dell’Universitá e della Ricerca under the program PRIN 2022 (“re-ranking of the final lists”), number 2022KWTEB7, cup B53C24006470006.

\section*{Author contributions}
\noindent
L.C. conceived the project. A.P. performed numerical simulations and analyzed data.  R.M., U.M.B.M., and L.C. wrote the first draft of the paper, but all authors contributed equally to manuscript writing.

\section*{Competing interests}
\noindent
The authors declare no competing interests.




\bibliographystyle{apsrev4-1}

\bibliography{odd.bib}

\end{document}